\documentclass[12pt]{iopart}
\usepackage{graphicx}
\usepackage{iopams}
\usepackage{bm}
\usepackage{dcolumn}
\def\rr{{{\bf r}}}

\begin{document}
\title[Equilibrium properties of simple metal thin films]{Equilibrium properties of simple metal thin films in the self-compressed stabilized jellium model}

\author{T Mahmoodi$^{1,2}$ and M Payami$^2$}

\address{$^1$ Department of Physics, Faculty of Sciences, Islamic Azad University, Mashhad Branch, Iran}

\address{$^2$ Physics Group, Nuclear Science and Technology Research Institute, Atomic Energy Organization of Iran, Tehran, Iran}

\ead{mpayami@aeoi.org.ir}

%
\begin{abstract}
In this work, we have applied the self-compressed stabilized jellium model to predict the equilibrium properties of isolated thin Al, Na, and Cs slabs. To make a direct correspondence to atomic slabs, we have considered only those $L$ values that correspond to $n$-layered atomic slabs with $2\le n\le 20$, for surface indices (100), (110), and (111).
The calculations are based on the density functional theory and self-consistent solution of the Kohn-Sham equations in the local density approximation. Our results show that firstly, the quantum size effects are significant for slabs with sizes smaller or near to the Fermi wavelength of the valence electrons $\lambda_{\rm F}$, and secondly, some slabs expand while others contract with respect to the bulk spacings. Based on the results, we propose a criterion for realization of significant quantum size effects that lead to expansion of some thin slabs. For more justification of the criterion, we have tested on Li slabs for $2\le n\le 6$. We have compared our Al results with those obtained from using all-electron or pseudo-potential first principles calculations. This comparison shows excellent agreements for Al(100) work functions, and qualitatively good agreements for the other work functions and surface energies. These agreements justify the way we have used the self-compressed stabilized jellium model for the correct description of the properties of the simple-metal slab systems. On the other hand, our results for the work functions and surface energies of large-$n$ slabs are in good agreement with those obtained from applying the stabilized jellium model for semi-infinite systems.  Moreover, we have performed the slab calculations in the presence of surface corrugation for a selected Al slabs and have shown that the results are worsened.
\end{abstract}
%

\maketitle
\section{\label{sec1}Introduction}
A thin slab is a system composed of a few atomic layers in such a way that it is finite in one direction (here we take as $z$ axis), and infinite in the other two directions of $x$ and $y$. The finiteness of the size in the $z$ direction gives rise to the so-called quantum-size effects (QSE)\cite{Schulte,Ho,Mola,Vicente,Boettger96,Boettger98,Kiejna99,Fall99,Sarria,DaSilva_05,Sferco,Zare}. The QSEs are significant when the size, $L$, of the system in one direction is comparable to the Fermi wavelength, $\lambda_{\rm F}$, of the electron in that direction and become less important in the limit $L\gtrsim 2\lambda_{\rm F}$.

To study the electronic structure and mechanical equilibrium properties of the simple metal slabs of Al, Na, and Cs, we have used the stabilized jellium model (SJM)\cite{Perdew90}. It has been shown that the SJM, in which the discrete ions of atoms are replaced by a uniform positive charge density, is a simple and realistic model to describe the bulk and surface properties of simple metals. Moreover, it has been shown that the self-compressed version of the SJM (SC-SJM) is appropriate for finite systems in which the surface effects causes the jellium background density differ from that of the bulk\cite{Perdew93,Payami_CJP04,Payami_prb06}. In the prior work by Sarria {\it et. al.}\cite{Sarria}, the SC-SJM has been applied to slabs with different sizes of $L$. For each fixed $L$ value, they have determined the $r_s^*$ value for the background that minimized the total energy per particle ($E/N$), and was always smaller than that of the bulk. That procedure was repeated for different values of $L$ parameter. They have shown that for certain values of $L$ which were integer multiples of $\lambda_{\rm F}/2$, the self-compression effects are more pronounced. However, since in their work the self-compression procedure was performed for constant $L$ values, any change of the $r_s$ value was equivalent to change in the number of atoms and thereby, the slab systems they studied, would not correspond to realistic isolated slabs. In order to apply the SC-SJM for isolated slabs, one has to consider $N$ as a fixed parameter and let the value of $L$ relax to its equilibrium value $L^\dagger$. For fixed $N$, the relaxation of $L$ relative to the corresponding bulk value leads to a change in the $r_s$ value. We denote by $r_s^\dagger$
the value corresponding to the $L^\dagger$, and will show in the following that under certain conditions it can assume values larger than that of the bulk.

In this work, using the SC-SJM, we have calculated the equilibrium properties of isolated thin slabs of simple
metals of Al, Na, and Cs. To make a direct correspondence to atomic slabs, we have considered only those $L$ values that correspond to $n$-layered atomic slabs with $2\le n\le 20$ for surface indices (100), (110), and (111). The calculations are based on the density functional theory (DFT)\cite{HK64} and solution of the Kohn-Sham (KS) equations\cite{KS65} in the local density approximation (LDA)\cite{KS65} for the
exchange-correlation (XC) functional. In our calculations, the slabs are taken as isolated
systems which can undergo relaxations in the $z$ direction. Our results show that, in contrast to the $r_s^*$ values (obtained for fictitious constant-volume slab system), for some slabs the $r_s^\dagger$ (obtained for realistic constant-$N$ slab system) assumes values that are larger than that of the bulk. That is, in some cases the relaxations are realized as expansions. The expansion behavior which is predicted in this work, have been also predicted by the other first-principles calculations\cite{Boettger98,DaSilva_05,Sferco,Zare}. The results show that the amount of relaxations decrease for large-$n$ slabs. Based on the results, we explore a criterion for realization of significant quantum size effects that lead to expansion of some thin slabs. For more justification of the criterion, we have tested on Li slabs for $2\le n\le 6$ and (in contrast to the results of Ref. \cite{Sarria}) have obtained some expansions. In section \ref{sec2} we have explained the
calculational details and in section \ref{sec3}, we have discussed on the results of our calculations and those obtained using first-principles methods\cite{Ho,DaSilva_05,Fall99,Kiejna99} as well as those obtained from SJM calculations for semi-infinite jellium systems\cite{Fiolhais,Kiejna93}. At the end, in Appendix, we have presented the detailed derivation of the relation between $r_s^*$ and $r_s^\dagger$.
\section{Calculational details}\label{sec2}
In the SC-SJM, the energy of a system with electronic charge distribution $n(\rr)$ and background charge $n_+(\rr)$ is given by\cite{Perdew93,PayamiJPC01}:
\begin{eqnarray}\label{eq1}
    E_{\rm SJM}[n,n_+]&=&E_{\rm JM}[n,n_+]+\left[\varepsilon_{\rm M}(r_s)+\bar{w}_R(r_s,r_c)       \right]\int d\rr n_+(\rr)\\ \nonumber
             &+& \langle\delta v\rangle_{\rm WS}(r_s,r_c)\int d\rr\;\Theta(\rr)[n(\rr)-n_+(\rr)],
\end{eqnarray}
where, $E_{\rm JM}$ is the energy in the jellium model (JM)\cite{Pay_Mah06}, $\varepsilon_{\rm M}=-9z/10r_0$ and $\bar{w}_R=2\pi\bar{n}r_c^2$ are the Madelung energy and the average of the repulsive part of the pseudopotential\cite{Ashcroft} over the Wigner-Seitz (WS) cell of radius $r_0=z^{1/3}r_s$, respectively. $z$ and  $\langle\delta v\rangle_{\rm WS}=(3r_c^2/2r_s^3-3z/10r_0)$ are the valency of the atom and the average of the difference potential over the WS cell, respectively\cite{Perdew90}.
 $n_+(\rr)=\bar{n}\Theta(\rr)$ is the jellium density in which $\bar n=3/4\pi r_s^3$ and $\Theta(\rr)$ takes the value of unity inside the jellium background and zero outside. $r_c$ is the core radius of the pseudopotential and is determined in such a way that the bulk system becomes mechanically stabilized at the experimental $r_s$ value. (All equations throughout this paper are expressed in hartree atomic units.)
 For a finite system (with fixed number of particles) at mechanical equilibrium, the $r_s$ value of the background is different from that of the bulk and assumes the value $r_s^\dagger$ for which
 \begin{equation}\label{eq2}
 \left.\frac{\partial(E/N)}{\partial r_s}\right|_{r_s^\dagger}=0,
 \end{equation}
where $N$ is the number of electrons in the system.

 In the SC-SJM, a slab of thickness $L$ has a full translational symmetry in the $x$ and $y$ directions, and therefore, the physical quantities depend on the spacial $z$ coordinate. Moreover, the KS equation reduces to a one-dimensional equation,
\begin{equation}\label{eq3}
    \left(-\frac{1}{2}\frac{d^2}{dz^2}+v_{eff}(z)\right)\psi_n(z)=\varepsilon_n\psi_n(z),
\end{equation}
where
\begin{equation}\label{eq4}
    v_{eff}(z)=\phi(z)+v_{xc}(z)+\langle\delta v\rangle_{\rm WS}\Theta(L/2-|z|).
\end{equation}
$\phi(z)$ is the electrostatic potential energy,
\begin{equation}\label{eq5}
    \phi(z)=4\pi\int_{-\infty}^z dz^\prime\,(z-z^\prime)[n_+(z^\prime)-n(z^\prime)],
\end{equation}
satisfying $\phi(\pm\infty)=0$, and the density is given by\cite{Schulte}:
 \begin{equation}\label{eq6}
    n(z)=\frac{1}{\pi}\sum_{\varepsilon_n\le E_{\rm F}}(E_{\rm F}-\varepsilon_n)\phi_n^2(z),
 \end{equation}
 with the Fermi energy determined by the charge neutrality condition.

To solve the KS equation for a slab of width $L$, we expand the KS orbitals in terms of the eigenfunctions of a  rectangular infinite potential well\cite{Mola} of width $D\approx L+4\lambda_{\rm F}$, which allows a $2\lambda_{\rm F}$ vacuum at each side of the slab. Our testings shows that this choice of the width $D$ ensures the independency of the results to the presence of the walls. The KS orbitals are expanded in terms of 150 lowest eigenfunctions of the rectangular box of width $D$. This number of basis ensures the reliability of energies to, at least, six significant figures.

For an $X(hkl)$ slab with $X$=Al, Na, Cs and $(hkl)$=(100), (110), (111), composed of $n$ atomic layers, the size is given by $L=nd_{hkl}$, with $d_{hkl}$ being the interlayer spacing. For a unit cell of $n_{at}$ atoms with valency $z$, the lattice constant $a$ is determined from the electronic $r_s$ value by $a=(4\pi z n_{at}/3)^{1/3}r_s$.
In Table~\ref{tab2} we have presented the numerical bulk values of the electronic density parameter $r_s$, lattice constant $a$ and the interlayer spacings $d_{hkl}$ for each element. It should be mentioned that since $\lambda_{\rm F}=2\pi(4/9\pi)^{1/3}r_s$ also linearly scales with $r_s$, the lattice constant in units of Fermi wavelength depends only on the valency of the atoms ($z$) and the crystal structure ($n_{at}$), {\it i.e.}, $a=(3zn_{at}/8\pi)^{1/3}\lambda_{\rm F}$.
\begin{table}
\caption{\label{tab2}Bulk $r_s$ values, lattice constants, and interlayer spacings in atomic units. The values in parentheses are in units of corresponding bulk Fermi wavelengths, $\lambda_{\rm F}$.}
\begin{indented}
\item[]\begin{tabular}{@{}ccccccc}
\br
  element & structure &$r_s$ & $a$ & $d_{100}$ & $d_{110}$ & $d_{111}$   \\
\mr
     Al   &fcc        & 2.07 & 7.64 (1.13) & 3.82 (0.56) & 2.70 (0.40)& 4.41 (0.65) \\
     Li   &bcc        & 3.28 & 6.60 (0.62) & 3.30 (0.31)& 4.66 (0.44)& 1.90 (0.18) \\
     Na   &bcc        & 3.99 & 8.10 (0.62) & 4.05 (0.31)& 5.73 (0.44)& 2.34 (0.18) \\
     Cs   &bcc        & 5.63 & 11.43 (0.62)& 5.72 (0.31)& 8.08 (0.44)& 3.30 (0.18) \\
\br
\end{tabular}
\end{indented}
\end{table}
To proceed with SC-SJM calculations for an isolated slab, we must use a reasonable relation between the $r_s$ and $L$ values. Since there is no relaxations in the $x$ and $y$ directions, the surface area of the slab remains constant and the change in $r_s$ is solely due to the change in $L$. Here, we assume the simplest relation satisfying the constant surface area and constant number of atoms,
\begin{equation}\label{eq7}
    L^\dagger=\left(\frac{r_s^\dagger}{r_s}\right)^3 L,
\end{equation}
where the pairs ($L^\dagger$, $r_s^\dagger$) and ($L$, $r_s$) correspond to the equilibrium and bulk states of the slab, respectively. Using this simple relation in our SC-SJM calculations, we predict that in the Na and Cs cases some slabs expand while, all Al slabs contract. Finally, to include the surface corrugation, we have used a face-dependent relation\cite{Perdew90}for $\langle\delta v\rangle$,
\begin{equation}\label{eq8}
    \langle\delta v\rangle_{\rm face}=\langle\delta v\rangle_{\rm WS}+\frac{z}{8r_0}\left\{\frac{12}{5}-\left[\frac{d}{r_0}\right]^2\right\},
\end{equation}
in which $d$ is the interlayer spacing for a given surface. For $d$ and $r_0$, we assume their corresponding bulk values. Our calculations show that the self-consistent variations of $d$ and $r_0$ do not lead to any equilibrium state. Moreover, eventhough the second term in the right hand side of Eq.~(\ref{eq8}) makes the work functions face-dependent, our results for Al(110) slabs with $n$=9, 11, 13, and 15 show that the results are worsened with respect to those obtained using a flat surface.

\begin{figure}
\begin{center}
\includegraphics[width=8.6cm,height=12cm]{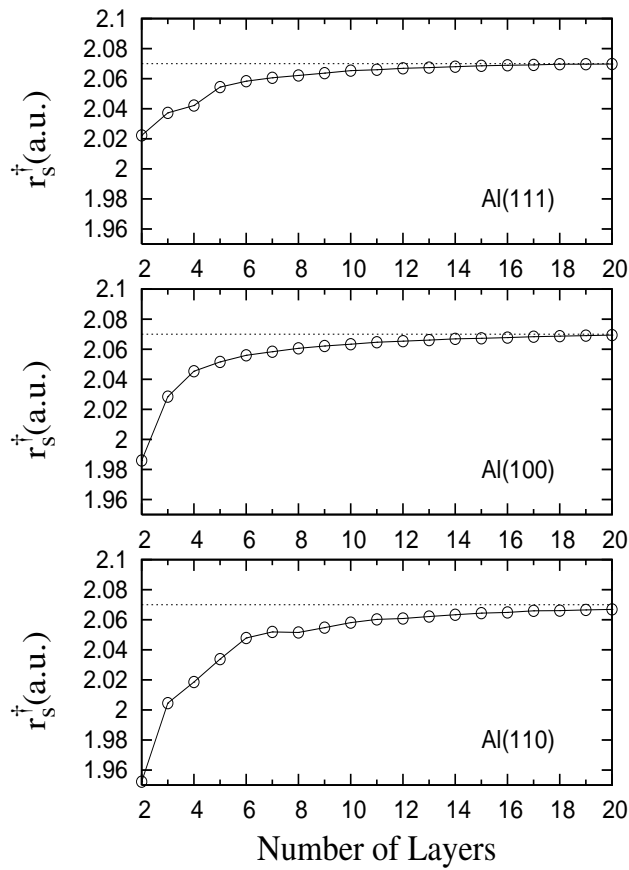}
\caption{\label{fig1}Equilibrium $r_s$ values, in atomic units, for $n$-layered Al slabs with (111), (100), and (110) surfaces. The dotted lines correspond to the bulk value (2.07).}
\end{center}
\end{figure}

\section{\label{sec3}Results and discussion}
Solving the self-consistent KS equations in the SC-SJM, we have calculated the equilibrium sizes, $L^\dagger$, of $n$-layered slabs ($2\le n\le 20$) for Al, Na, and Cs. For each element, the $n$-layered slab calculations has been repeated for the different (100), (110), and (111) surface indices. Here, it is assumed that the interlayer spacing is the only parameter that differs for differet surface indices. However, as we discuss later in this section, by introducing the surface corrugation, $\langle\delta v\rangle$ becomes also face-dependent which leads to different surface properties   for slabs of different indices. The advantages of using $d_{hkl}$ is that we consider only those $L$ values that correspond to real atomic slabs and therefore, for each structure there is a one-to-one correspondence to atomic slabs. For each $n$-layered slab, the KS equations are solved self-consistently for different $L$ values and thereby, the value $L^\dagger$ which minimizes the total energy per particle, is determined.

In Fig.~\ref{fig1}, we have plotted the $r_s^\dagger$ for Al(111), Al(100), and Al(110) as function of the number of atomic layers, $n$. (Throughout the paper, all sub-figures from top to bottom, are ordered according to a decreasing bulk interlayer spacing.) The results show that $r_s^\dagger<2.07$ for all Al slabs. That is, the model predicts all Al slabs are contracted. Comparison of the three sub-figures in Fig.~\ref{fig1} and taking into account the different interlayer spacings from Table \ref{tab2}, we see that the smaller the size of the slabs the larger is the contraction rate. Moreover, since the sizes of the 2-layered Al slabs are larger than $\lambda_{\rm F}$, we do not observe any significant oscillations.

\begin{figure}
\begin{center}
\includegraphics[width=8.6cm,height=12cm]{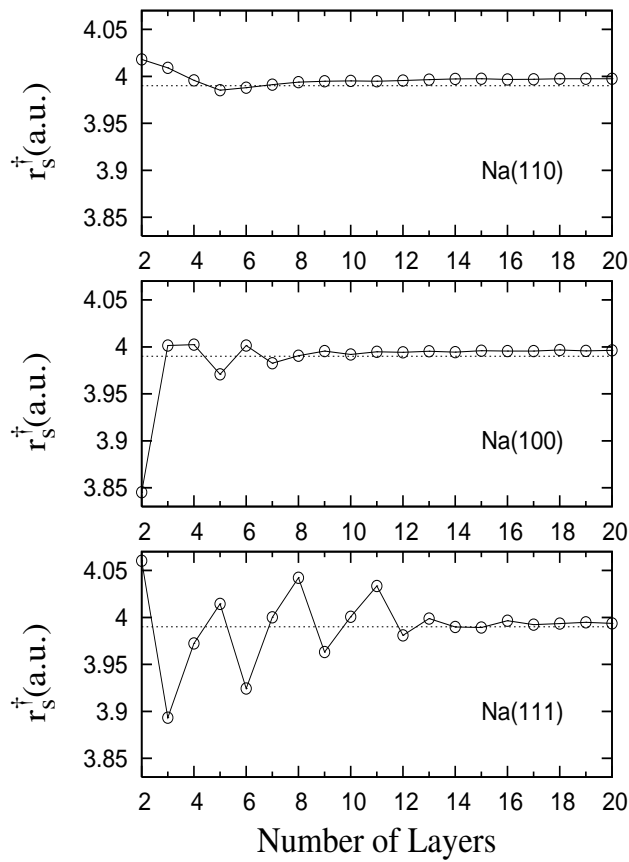}
\caption{\label{fig2}Same as in Fig.~\ref{fig1} for Na, with bulk value 3.99.}
\end{center}
\end{figure}

In Fig.~\ref{fig2}, we have plotted the $r_s^\dagger$ results for Na, which shows significant deviations from the bulk $r_s$ value up to $n$=13 for Na(111), up to $n$=8 for Na(100), and up to $n$=5 for Na(110). These sizes correspond to slab thicknesses of $\sim 2\lambda_{\rm F}$. For sizes above these thresholds the oscillations become negligible. That is, the QSEs are significant for $L\lesssim 2\lambda_{\rm F}$. On the other hand, we observe that the oscillations become sharper and significant as we go from top subfigure to the bottom. Since from top to bottom the interlayer spacing decreases, we are led to the conclusion that if the interlayer spacing, $d$, happens to be so small or the Fermi wavelength happens to be so large that the condition $2\lambda_{\rm F}/d \gg 1$ is met, then there would be a chance that the QSE be realized as consecutive expansions and contractions. However, since this condition is not met in the Al case ($d\sim 0.5\lambda_{\rm F}$), we do not observe any significant oscillations. This criterion along with the first-principles or experimental results enables one to estimate an effective free-electron $r_s$ value for a given slab. For example, as is shown in the following (Fig.~\ref{fig4}), according to first-principles calculations, the Al(100) and Al(111) slabs undergo expansions\cite{DaSilva_05}. This means that, using the criterion, the actual Fermi wavelengths along those directions should be larger than the bulk free-electron value or the actual electron density in the mid of two adjacent atomic layers should be much smaller than that of the bulk.

\begin{figure}
\begin{center}
\includegraphics[width=8.6cm,height=12cm]{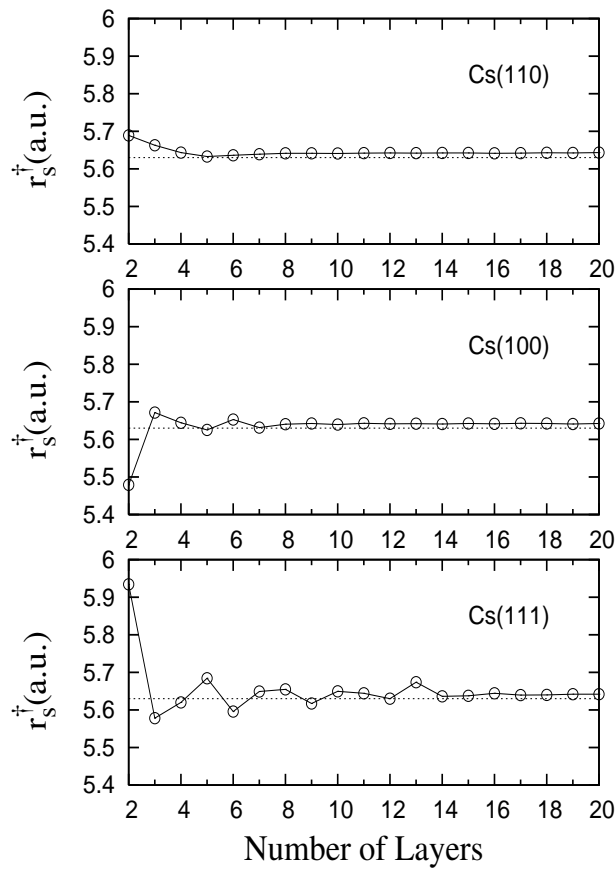}
\caption{\label{fig3}Same as in Fig.~\ref{fig1} for Cs, with bulk value 5.63.}
\end{center}
\end{figure}

The $r_s^\dagger$ results for Cs slabs are shown in Fig.~\ref{fig3}. As is observed, the shapes are similar to the corresponding shapes of Na slabs. This fact is explained by noticing that in Cs the interlayer spacings, in units of $\lambda_{\rm F}$, are the same as those in Na (see Table~\ref{tab2}). On the other hand, since the density parameter, $r_s$, of the electrons in Cs is larger than that in Na, the oscillations in Cs case have smaller amplitudes relative to those in Na.

\begin{figure}
\begin{center}
\includegraphics[width=8.6cm,height=12cm]{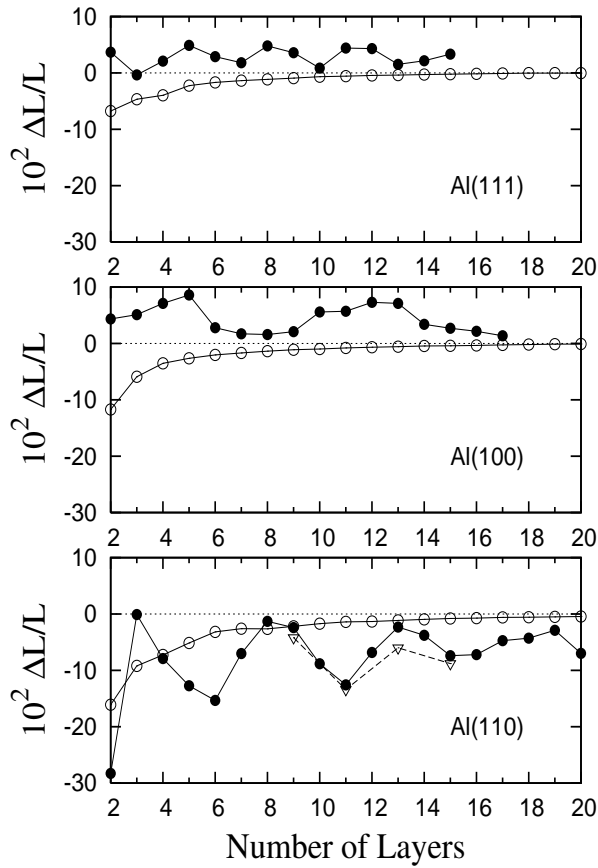}
\caption{\label{fig4}Relative total relaxations of Al slabs at their equilibrium states for (111), (100), and (110) surfaces. The open circles, solid circles, and upside-down triangles correspond to the results of present work, FPLAPW\cite{DaSilva_05}, and PP-LDA\cite{Ho}, respectively.}
\end{center}
\end{figure}

According to Eq.~(\ref{eq7}), the variations in $L$ and $r_s$ are related by
\begin{equation}\label{eq20}
    \frac{\Delta r_s}{r_s}=\left(\frac{\Delta L}{L} \right)^{1/3}-1,
\end{equation}
which can be used to determine the total thickness relaxations of the slabs. In Fig.~\ref{fig4}, we have compared the results of total relaxations of the Al slabs with those of first-principles calculations\cite{DaSilva_05,Ho}. In the top sub-figure, we have compared the results on Al(111) with those obtained\cite{DaSilva_05} from full-potential linearized augmented plane wave method (FPLAPW) in the generalized gradient approximation (GGA)\cite{GGA}. The FPLAPW results show expansions up to $n$=17 whereas our results show contractions. As we have argued, this could be as a result of larger actual electron Fermi wavelength in that direction compared to that of the bulk. Similarly, in the middle sub-figure, the FPLAPW results show expansions, while our results show contractions. However, the (110) slabs in the bottom subfigure show contractions as predicted by FPLAPW and LDA pseudo-potential (PP) calculations\cite{Ho}. In this case, the average behavior of the first-principles results agrees well with our results, and we can conclude that $\lambda_{\rm F}$ along this direction is not much different from that in the free-electron model.

\begin{figure}
\begin{center}
\includegraphics[width=8.6cm,height=12cm]{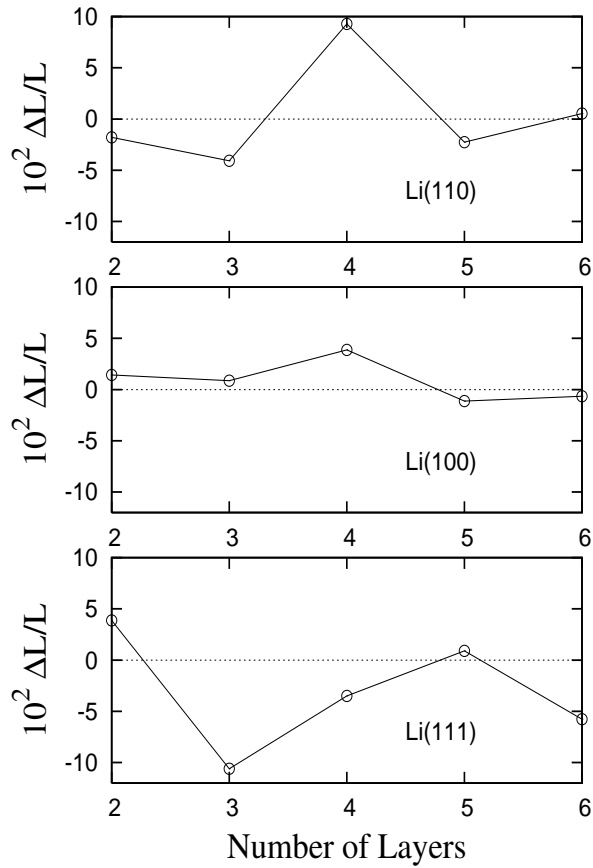}
\caption{\label{fig5}Same as in Fig.~\ref{fig4} for Li.}
\end{center}
\end{figure}

The results of total relaxations for Na and Cs behave similar to their corresponding $r_s^\dagger$ plots and are not presented. To close the discussion on the thickness relaxations, we have presented the relaxations of Li slabs for $n$=2, 3, 4, 5, 6 in Fig.~\ref{fig5}. The results show expansions at $n$=4, 6 for Li(110); $n$=2, 3, 4 for Li(100);  and $n$=2, 5 for Li(111), respectively. These expansions are verifications of our criterion for observing significant QSEs. In addition, by Eq.~(\ref{eq20}), at these $n$ values, the $r_s^\dagger$s are greater than that of the bulk while the $r_s^*$ values, obtained for Li in Ref.~\cite{Sarria}, were always smaller than that of the bulk.

\begin{figure}
\begin{center}
\includegraphics[width=8.6cm,height=12cm]{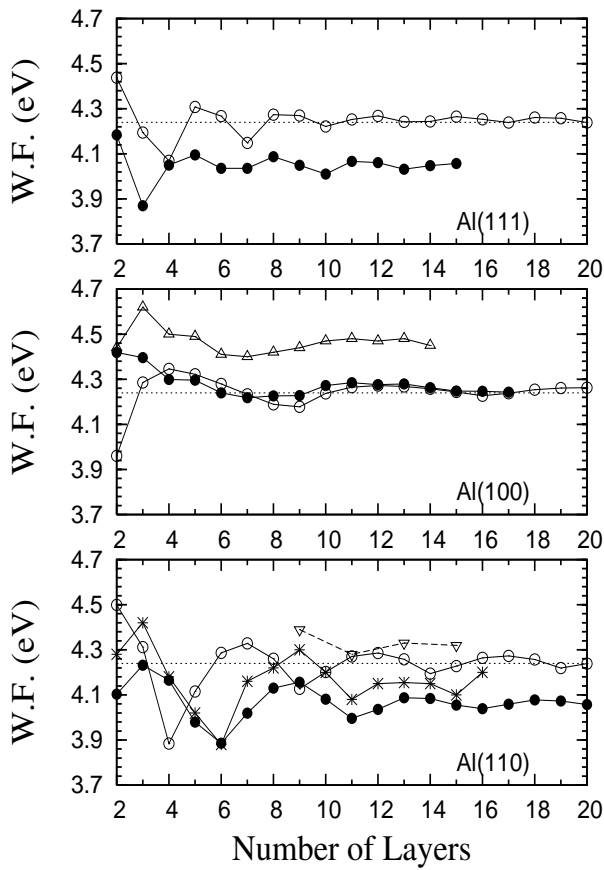}
\caption{\label{fig6} Work functions of Al slabs at their equilibrium states for (111), (100), and (110) surfaces as functions of number of layers. The open circles, solid circles, open triangles, asteriscs, and upside-down open triangles correspond to the results of present work, FPLAPW\cite{DaSilva_05}, PP-LDA\cite{Fall99}, PP-LDA\cite{Kiejna99}, and PP-LDA\cite{Ho}, respectively. The dotted lines correspond to the SJM results for the semi-infinite system\cite{Fiolhais,Kiejna93}. }
\end{center}
\end{figure}

The work function of a given metal surface is the minimum energy needed to remove an electron from inside of the metal to a point outside across that surface. In the slab system, it is equal to the absolute value of the Fermi energy. In Fig.~\ref{fig6}, we have plotted the work function results of Al slabs and compared with those of FPLAPW\cite{DaSilva_05}. However, for completeness we have also included the pseudo-potential calculation results in the LDA (PP-LDA)\cite{Ho,Fall99,Kiejna99} as well as the SJM results for semi-infinite system\cite{Fiolhais,Kiejna93}. In order to explain our results, as is observed from FPLAPW results, in Al(100), there is an excellent agreement in the values and oscillation behavior for $n\ge 4$. On the other hand, the agreement is poor for $n$=2, 3. In other words, the bending of the FPLAPW curve is a little bit slower than that in our results. Inspecting the top and bottom sub-figures also shows that our results and those of FPLAPW have similar oscillatory structures. The relative vertical shifts at asymptotic large-$n$ regions can be related to the differences in the values of the surface dipole moments (SDM). The SDMs in our model calculations are higher than those in the FPLAPW or PP calculations. It is because, in the SC-SJM the background density has the same constant value over the slab width, whereas in the FPLAPW or PP the discrete ions of each plane are allowed to find their own equilibrium positions. It means that, if we had taken an inhomogeneous jellium background to simulate the discrete ionic planes, then the equilibrium densities of background sub-slabs would be different. This behavior will somehow decrease the separation between the positive and negative charges and thereby decreases the SDM. In the extreme case of complete deformable background, the ultimate jellium model\cite{Koskinen}, there is no charge separations and therefore, the SDM vanishes. On the other hand, we should not forget that in the FPLAPW calculations the GGA has been used which gives results different from that in the LDA for highly inhomogeneous electron densities at surfaces. This fact is shown by presenting the PP-LDA results in the middle and bottom sub-figures. The PP results lie above those of the FPLAPW with the same structure.
Returning back to the oscillatory structure, the local Fermi wavelengths of the electrons in the FPLAPW are different from those of SC-SJM. Under the condition $(\lambda_{\rm F}/d)\gtrsim 1$, a small decrease in $\lambda_{\rm F}$ (other parameters kept fixed) results in a decrease of the value of the ratio, and in turn, an increase in the bending rate. It is the case in our Al(100) results. Here, slabs with $n$=2, 3 have been contracted while those in the FPLAPW are expanded (See middle sub-figure in Fig.~\ref{fig4}). It should be mentioned that for small $n$'s no bulk region exists for the slab (unless $\lambda_{\rm F}/d\ll 1$), and expansion or contractions will have different impacts on the surface properties. However, for large $n$'s, the effects of relaxations in the bulk can be canceled out by the relaxations of the neighboring layers and therefore, the SDM does not change. In other words, the relaxation of the layers which lie inside the major peak of the Friedel oscillation, determines the value of the SDM. That is why there is an excellent agreement between SC-SJM and FPLAPW work functions for large $n$, in spite of the fact that our Al(100) slabs contract while Al(100) slabs expand in FPLAPW.
To explain the differences between our results and those of FPLAPW in the work function of Al(110) slabs, we notice from the bottom sub-figure of Fig.~\ref{fig4} that FPLAPW predicts more compression than SC-SJM for $n$=2, and hence, smaller Fermi wavelength which leads to a rapid bending of the work function curve. For $n>4$, we notice a constant phase difference between the two oscillatory curves. That is, (aside from the vertical shift which we discussed) the value of the SDM at $n$=4 in SC-SJM is equivalent to that in FPLAPW at $n$=6, and so on. This fact can be explained by assuming that the overlap between the major peaks in the  Friedel oscillation at both sides of the slab effectively vanishes for $n\ge 6$ atomic layers in the FPLAPW and for $n\ge 4$ in the SC-SJM. For larger $n$'s, the constant phase difference means that the $\lambda_{\rm F}$ of the electrons in the bulk region is the same in the SC-SJM and the FPLAPW.
Now, we focus on the top sub-figure of Fig.~\ref{fig6} on Al(111). For $n\ge 7$, aside from the vertical shift, the two curves have the same frequency of oscillations and therefore, in the bulk region both SC-SJM and FPLAPW have the same electronic $\lambda_{\rm F}$. However, for $n<7$, from $n$=2 to $n$=3 FPLAPW gives contraction while from $n$=3 to $n$=7 it gives expansion (See top sub-figure in Fig.~\ref{fig4}). Therefore, going from $n$=2 to $n$=7, first $\lambda_{\rm F}$ decreases and then increases. As a result, in the FPLAPW work function plot we observe a rapid bending followed by a slow bending. From the FPLAPW results, it can be deduced that at $n$=7, the surface formation is completed. To sum up, (from Fig.~\ref{fig4}) FPLAPW results show positive slopes at $n$=2, followed by a negative one
for Al(100) and Al(110), whereas it shows negative slope at $n$=2 followed by a positive one for Al(111). These behaviors are reflected in the work function curves (Fig.~\ref{fig6}) as slow bending at $n$=2 for Al(100), Al(110), and rapid bending for Al(111).
The dotted straight lines in each sub-figures correspond to the result obtained from the self-consistent SJM calculations for semi-infinite system\cite{Fiolhais,Kiejna93}. For completeness, we have also included the PP-LDA results\cite{Ho,Fall99,Kiejna99}.

To see the effects of surface corrugation on the work functions, we have used the face-dependent relation of Eq.~(\ref{eq8}) for $\langle\delta v\rangle$ for a selected Al(110) slabs. The results are presented in Table~\ref{tab4}.

\begin{table}[hb]
\caption{\label{tab4}Comparison of the relaxations, averaged difference potentials, and work functions of $n$-layered Al slabs as well as semi-infinite jellium results with those of FPLAPW.}
\footnotesize\rm
\begin{tabular}{@{}cc|ccc|ccc|cc}
\br
 $n$ & case & $\Delta L/L(\%)^{\rm a}$ & $\langle\delta v\rangle$~(eV)$^{\rm a}$& W~(eV)$^{\rm a}$& $\Delta L/L(\%)^{\rm b}$ & $\langle\delta v\rangle$~(eV)$^{\rm b}$& W~(eV)$^{\rm b}$&
 $\Delta L/L(\%)^{\rm c}$ &W~(eV)$^{\rm c}$ \\
\mr
 9  & (100)& -1.12 & -2.45 & 4.18 & -2.67 & 0.20 & 3.69 &  2.1  &   4.23   \\
    & (110)& -2.19 & -2.42 & 4.13 & -6.55 & 3.13 & 3.56 & -2.4  &   4.16   \\
    & (111)& -0.91 & -2.46 & 4.27 & -1.24 & -1.70& 4.09 &  3.6  &   4.05   \\
 11 & (100)& -0.78 & -2.46 & 4.26 & -2.02 & 0.18 & 3.78 &  5.7  &   4.28   \\
    & (110)& -1.41 & -2.44 & 4.27 & -5.32 & 3.09 & 3.61 & -12.6 &   4.00   \\
    & (111)& -0.58 & -2.47 & 4.25 & -0.86 & -1.71& 4.07 &  4.4  &   4.07   \\
 13 & (100)& -0.56 & -2.47 & 4.27 & -1.56 & 0.17 & 3.77 &  7.1  &   4.28   \\
    & (110)& -1.13 & -2.45 & 4.26 & -4.44 & 3.06 & 3.62 & -2.3  &   4.09   \\
    & (111)& -0.38 & -2.47 & 4.24 & -0.61 & -1.72& 4.06 &  1.5  &   4.03   \\
 15 & (100)& -0.40 & -2.47 & 4.24 & -1.25 & 0.16 & 3.74 &  2.7  &   4.25   \\
    & (110)& -0.81 & -2.46 & 4.23 & -3.64 & 3.03 & 3.54 & -7.4  &   4.06   \\
    & (111)& -2.08 & -2.48 & 4.26 & -0.41 & -1.73& 4.08 &  3.3  &   4.06   \\
 20 & (100)& -0.09 & -2.48 & 4.26 & -0.07 & 0.14 & 3.76 &  --   &   --     \\
    & (110)& -0.44 & -2.47 & 4.24 & -2.44 & 3.00 & 3.55 &  --   &   --     \\
    & (111)& -0.02 & -2.48 & 4.24 & -0.01 & -1.74& 4.05 &  --   &   --     \\
\mr
semi-inf$^{\rm d}$& flat & -- & -2.49& 4.24  & --  & -- & --  & -- &--    \\
                   & (100)& -- & --   & -- & -- &   0.12 & 3.62  &-- &--  \\
                   & (110)& -- & --   & -- & -- &   2.92 & 3.81  & --&--  \\
                   & (111)& -- & --   & -- & -- &   -1.74& 3.72  & --&--  \\
\br
\end{tabular}
$^{\rm a}$ {Present work, flat surface.}\\
$^{\rm b}$ {Present work, face-dependent corrugation.}\\
$^{\rm c}$ {FPLAPW (Ref~\cite{DaSilva_05}).}\\
$^{\rm d}$ {Semi-infinite SJM system from Ref.~\cite{Fiolhais}. This result is quite close to that in Ref.~\cite{Kiejna93} which is 4.27 eV.}
\end{table}

In Table~\ref{tab4}, the numbers in the fifth column are the work functions of flat surface slabs. The eighth and tenth columns are the face-dependent and the FPLAPW results, respectively. As is observed from the Table~\ref{tab4}, the surface corrugation term has affected the amount of relaxation and the averaged potential difference, $\langle\delta v\rangle$. Any change in the difference potential, causes a change in the SDM, and thereby lead to a change in the work function. However, the (100) work functions of flat surface agrees nicely with those of FPLAPW\cite{DaSilva_05}, and therefore, the face-dependent term worsens the results. On the other hand, work functions of both (110) and (111) slabs with flat surfaces are higher than those of FPLAPW, but, the corrugation term causes overcorrection and worsening the agreement with FPLAPW. The lowest four rows of the table show the results obtained from application of the SJM to semi-infinite jellium system\cite{Fiolhais} which are in excellent agreement with our results for large $n$'s.

\begin{figure}
\begin{center}
\includegraphics[width=8.6cm,height=12cm]{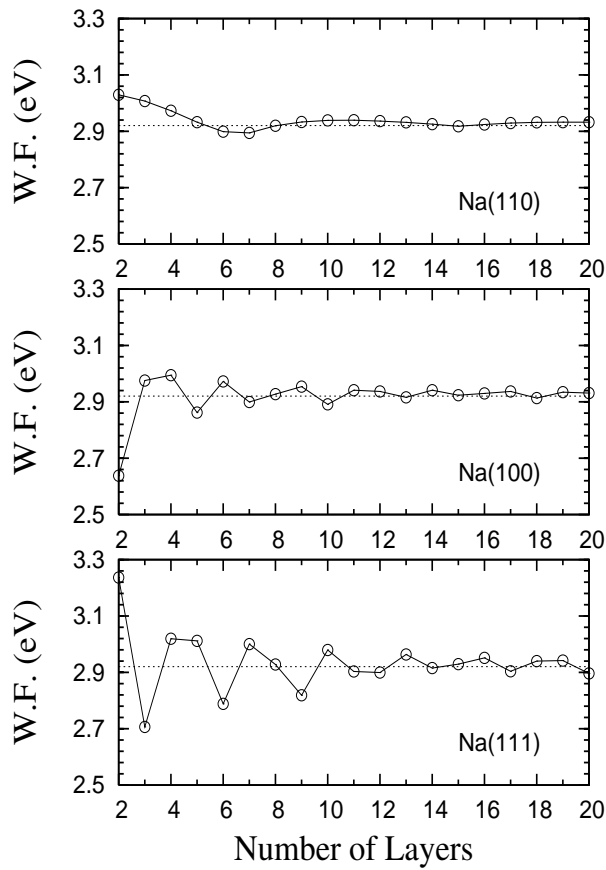}
\caption{\label{fig7}Same as in Fig.~\ref{fig6} with semi-infinite results 2.92~eV\cite{Fiolhais} which is close to 2.94~eV\cite{Kiejna93}.}
\end{center}
\end{figure}

\begin{figure}
\begin{center}
\includegraphics[width=8.6cm,height=12cm]{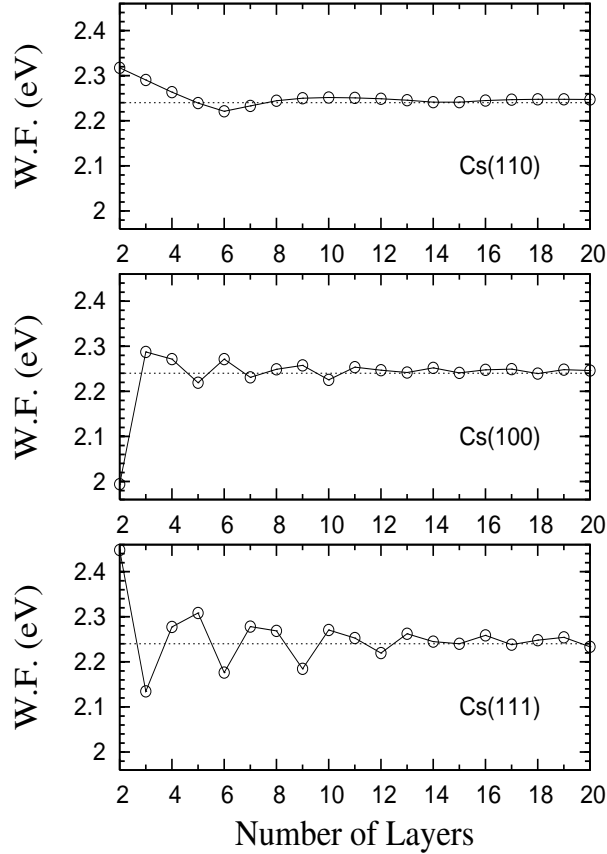}
\caption{\label{fig8}Same as in Fig.~\ref{fig6} with semi-infinite results 2.24~eV\cite{Fiolhais} which is close to 2.26~eV\cite{Kiejna93}.}
\end{center}
\end{figure}

In Fig.~\ref{fig7} we have presented the work functions of Na slabs at their equilibrium states for (110), (100), and (111) surfaces as functions of number of layers. The dotted lines correspond to the SJM results (2.92~eV) for semi-infinite system\cite{Fiolhais} which is close to 2.94~eV obtained in Ref.~\cite{Kiejna93}.

The calculation results for the work functions of Cs slabs are plotted in Fig.~\ref{fig8}.
The dotted lines correspond to the SJM results (2.24~eV) for semi-infinite system\cite{Fiolhais} which is close to 2.26~eV obtained in Ref.~\cite{Kiejna93}. As in the case of $r_s^\dagger$ results, here also, the behaviors are similar to those of corresponding Na slabs.

\begin{figure}
\begin{center}
\includegraphics[width=8.6cm,height=12cm]{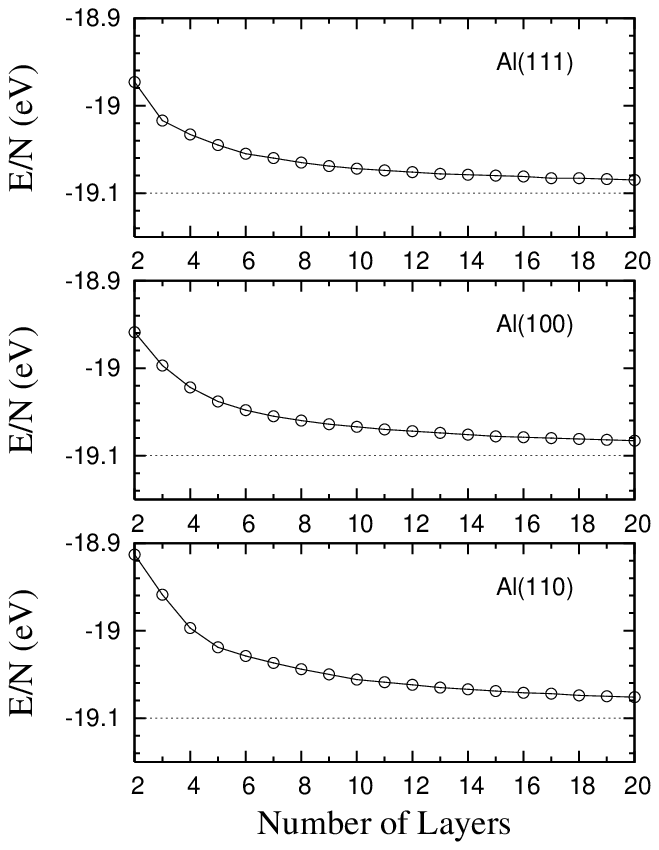}
\caption{\label{fig9}Total energies per electron of Al slabs at their equilibrium states for (111), (100), and (110) surfaces. The dotted lines correspond to the bulk value, -19.1~eV \cite{Perdew90}.}
\end{center}
\end{figure}

In Fig.~\ref{fig9}, we have plotted the total energies per electron for Al(111), Al(100), and Al(110) slabs at their equilibrium states. As seen, the behaviors are quite smooth and approach to the bulk value -19.1~eV \cite{Perdew90} from above for large enough $L$'s. Similar behaviors are obtained for Na and Cs slabs with respective asymptotic values of -6.26 and -4.64~eV \cite{Perdew90}.

\begin{figure}
\begin{center}
\includegraphics[width=8.6cm,height=12cm]{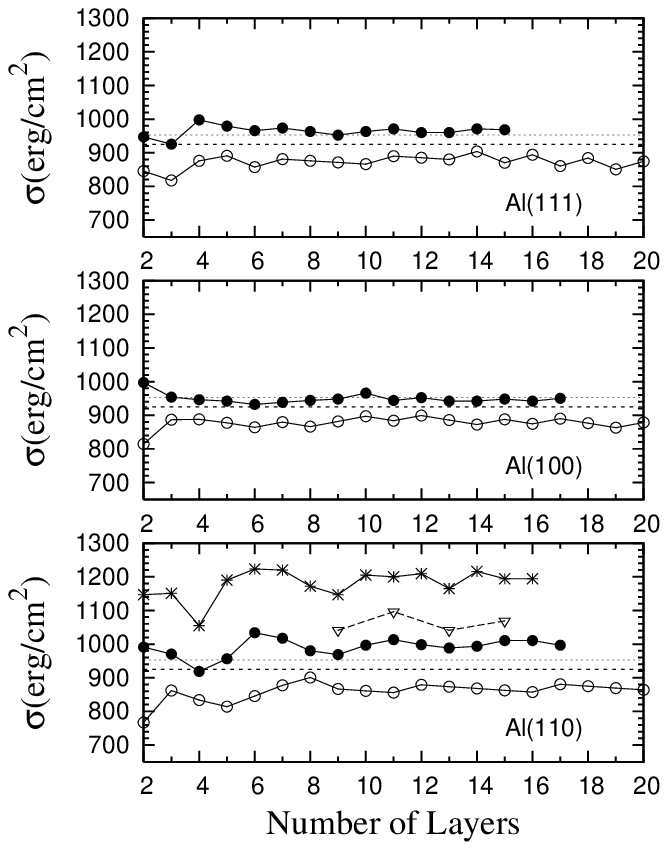}
\caption{\label{fig10}Surface energies of Al slabs at their equilibrium states for (111), (100), and (110) surfaces. The labels are the same as in Fig.~\ref{fig6}. The dotted and dashed lines correspond to the SJM results for the semi-infinite system with 953~erg/cm$^2$\cite{Fiolhais} and 925~erg/cm$^2$\cite{Kiejna93}, respectively.}
\end{center}
\end{figure}

\begin{figure}
\begin{center}
\includegraphics[width=8.6cm,height=12cm]{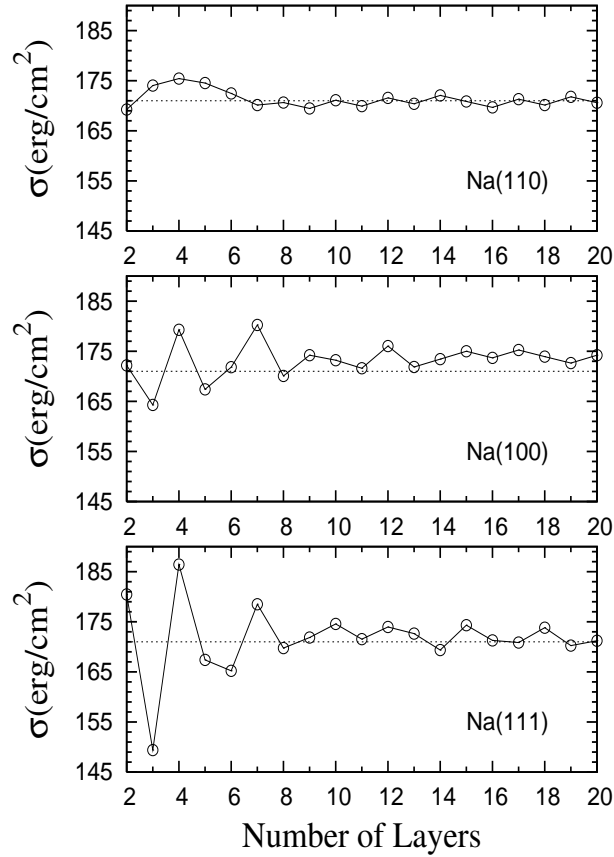}
\caption{\label{fig11}Same as in Fig.~\ref{fig10} for Na with 171~erg/cm$^2$ for semi-infinite SJM\cite{Fiolhais,Kiejna93}.}
\end{center}
\end{figure}

\begin{figure}
\begin{center}
\includegraphics[width=8.6cm,height=12cm]{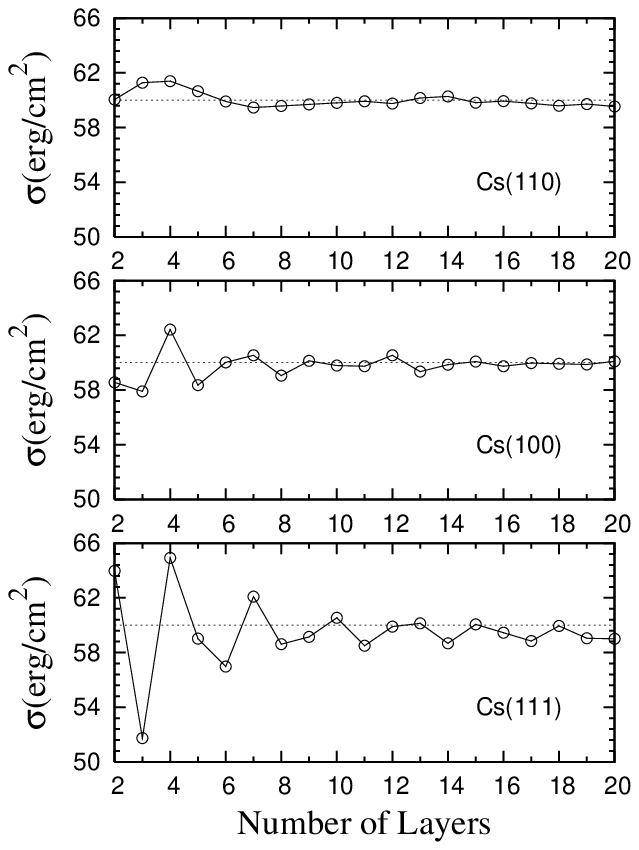}
\caption{\label{fig12}Same as in Fig.~\ref{fig10} for Cs with 60~erg/cm$^2$ for semi-infinite SJM\cite{Fiolhais,Kiejna93}.}
\end{center}
\end{figure}

The surface energy of the slab is defined by
\begin{equation}\label{eq21}
    \sigma(L^\dagger)=\frac{1}{2A}\left[E_{\rm SJM}(L^\dagger)-E_{\rm SJM}^{bulk}(L) \right],
\end{equation}
where $E_{\rm SJM}(L^\dagger)$ and $E_{\rm SJM}^{bulk}(L)$ are the SC-SJM  and bulk-SJM total energies of the slab with surface area $A$. The bulk-SJM total energy is proportional to $L=nd_{hkl}$ and Eq.~(\ref{eq21}) can be rewritten in the form of
\begin{equation}\label{eq22}
    \frac{E_{\rm SJM}(L^\dagger)}{A}=2\sigma(L^\dagger)+L\bar n \varepsilon^{bulk},
\end{equation}
where $\bar n$ and $\varepsilon^{bulk}$ are the bulk density and bulk SJM energy per particle, respectively. We have fitted the values of the SC-SJM total energy per unit area to Eq.~(\ref{eq22}) to obtain $\varepsilon^{bulk}$, and used it in Eq.~(\ref{eq21}) to obtain $\sigma(L^\dagger)$. In the fitting we have used the results with $n\ge 6$ to eliminate the significant fluctuations and obtain more accurate values for $\varepsilon^{bulk}$.

In Fig.~\ref{fig10}, we have plotted the surface energies obtained from using Eq.~(\ref{eq21}) for Al slabs with flat surfaces. The results are compared with those obtained from the FPLAPW\cite{DaSilva_05}. Comparison shows a systematic underestimation of $\sim 8\%$ for Al(111), $\sim 9\%$ for Al(100), and $\sim 10\%$ for Al(110) surfaces, respectively, which can be partly due to the GGA used in the FPLAPW calculations. In the bottom sub-figure, Al(110), we have also included the PP-LDA results which lie above both the SC-SJM and FPLAPW results. As is observed, the structures of the plots are in excellent agreement with those of FPLAPW. On the other hand, it is seen that both PP and FPLAPW plots have the same structures. The disagreements of $n$=2 in Al(100) and in Al(110) with those in FPLAPW are consistent with those in Fig.~\ref{fig6}. The phase difference is also observed in the bottom sub-figure. The dotted and dashed lines correspond to the results of SJM for semi-infinite system with 953~erg/cm$^2$\cite{Fiolhais} and 925~erg/cm$^2$\cite{Kiejna93}, respectively.

The surface energies of Na and Cs slabs are plotted in Figs.~\ref{fig11} and \ref{fig12}, respectively. Consistent with the top sub-figures in Figs.~\ref{fig2}, \ref{fig3}, \ref{fig7}, \ref{fig8}, the top sub-figures here also show smooth behaviors. Compared to any FPLAPW calculations, we expect a systematic underestimation of surface energies of at most $\sim 10\%$ for Na and Cs as well. On the other hand, as in Al case, our results for Na and Cs are in excellent agreement with those of the semi-infinite SJM for flat surfaces\cite{Fiolhais,Kiejna93}. Finally, we note that the semi-infinite calculations\cite{Perdew90} with surface corrugations have raised the Al surface energies to 977, 1103, and 921 erg/cm$^2$ for (100), (110), and (111) indices, respectively, which are in better agreements with those of FPLAPW.

\section{\label{sec4}Conclusions}
We have applied the SC-SJM to predict the electronic properties of simple metal slabs of Al, Na, and Cs at their equilibrium states. To make a direct correspondence to atomic slabs, we have considered only those $L$ values that correspond to $n$-layered atomic slabs with $2\le n\le 20$ for surface indices (100), (110), and (111).
In our calculations, the slabs are taken as isolated
systems which can undergo relaxations in the $z$ direction. Our results show that, in contrast to the $r_s^*$ values (obtained in Ref.~\cite{Sarria}) which are always smaller than those of the bulk, for some slabs the $r_s^\dagger$'s assume values that are larger than those of the bulk. That is, in some cases the relaxations are realized as expansions.
Our results show that all Al slabs are contracted while some of Na and Cs slabs expand. From these results we explore that if the interlayer spacing happens to be so small or the Fermi wavelength happens to be so large that the condition $2\lambda_{\rm F}/d \gg 1$ is met, then there would be a chance that the QSE be realized as consecutive expansions and contractions. Since this condition is not met in the Al case, we do not see any significant sharp oscillations in the QSEs.
For more justification of the criterion, we have tested on Li slabs for $2\le n\le 6$ and have obtained some expansions.  This criterion along with the first-principles or experimental results can be used to estimate an effective free-electron $r_s$ value for a given slab. On the other hand, our results for the work functions and surface energies show similar structures as in those of the FPLAPW results. We have explained the quantitative differences between our results and those of the FPLAPW to be partly due to the fact that our model calculations overestimate the SDM values, and partly because of the fact that in FPLAPW calculations, the GGA has been used which give significant improvements over the LDA results for systems with high inhomogeneities in electron densities. Finally, our results are in excellent agreement with the self-consistent results of the SJM for semi-infinite system.
\ack{
MP would like to appreciate professor John~P.~Perdew for his useful comments. This work is part of research program in NSTRI, Atomic Energy Organization of Iran.}
%
\appendix
\section{Relation between $r_s^\dagger$ and $r_s^*$}
The energy of the slab system is a function of volume, $V$, and $r_s$, {\it i.~e.,} $E=E(V,r_s)$. In fact, instead of $V$, it depends on $L$ because, in the slab system the surface area does not change. However, we use $V$ for generality of our arguments. We define two different derivatives, $D_N$ and $D_V$, which are derivatives when $N$ is constant and when $V$ is constant, respectively. That is,
\begin{eqnarray}\label{eq9}
   \nonumber D_N &\equiv& \frac{d}{dr_s}\left[\frac{E(V,r_s)}{N}\right]_N = \frac{1}{N}\frac{d}{dr_s}\left[E(V,r_s)\right]_{N}\\
   & =& \frac{4\pi}{3V}r_s^3\frac{d}{dr_s}\left[E(V,r_s) \right]_{N},
\end{eqnarray}
\begin{eqnarray}\label{eq10}
  \nonumber  D_V &\equiv & \frac{d}{dr_s}\left[\frac{E(V,r_s)}{N}\right]_V = \frac{1}{V}\frac{d}{dr_s}\left[\frac{V}{N}E(V,r_s) \right]_{V}\\
   &=& \frac{4\pi}{3V}\frac{d}{dr_s}\left[r_s^3E(V,r_s) \right]_{V}.
\end{eqnarray}
However, on the one hand, using Eq.~(\ref{eq7}) one obtains
\begin{equation}\label{eq11}
    \frac{dV}{dr_s}=\frac{3V}{r_s},
\end{equation}
which results in
\begin{eqnarray}\label{eq12}
  \frac{d}{dr_s}\left[E(V,r_s)\right]_N &=& \left[\left(\frac{\partial E}{\partial V} \right)_{r_s}\frac{dV}{dr_s}+
  \left(\frac{\partial E}{\partial r_s} \right)_V \right]_N \\ \nonumber
   &=& \frac{3}{r_s}V\left(\frac{\partial E}{\partial V} \right)_{r_s}+
  \left(\frac{\partial E}{\partial r_s} \right)_V,
\end{eqnarray}
and on the other hand, the constant volume conditions results in
\begin{equation}\label{eq13}
    \frac{d}{dr_s}\left[E(V,r_s)\right]_V=\left(\frac{\partial E}{\partial r_s} \right)_V.
\end{equation}
Now, combining Eqs.~(\ref{eq9})-(\ref{eq13}) one obtains
\begin{equation}\label{eq14}
D_V(V,r_s)=D_N(V,r_s)+\frac{4\pi r_s^2}{V}\left[ E(V,r_s)-V\left(\frac{\partial E}{\partial V} \right)_{r_s} \right]
\end{equation}
By definition,
\begin{equation}\label{eq15}
    D_V(V,r_s^*)=0,
\end{equation}
and
\begin{equation}\label{eq16}
    D_N(V,r_s^\dagger)=0.
\end{equation}
Inserting Eqs.~(\ref{eq15}) and (\ref{eq16}) into (\ref{eq14}) results in
\begin{equation}\label{eq17}
    D_N(V,r_s^*)+\frac{4\pi r_s^{* 2}}{V}\left[ E(V,r_s^*)-V\left(\frac{\partial E}{\partial V} \right)_{r_s^*} \right]=0,
\end{equation}
and
\begin{equation}\label{eq18}
    D_V(V,r_s^\dagger)-\frac{4\pi r_s^{\dagger 2}}{V}\left[ E(V,r_s^\dagger)-V\left(\frac{\partial E}{\partial V} \right)_{r_s^\dagger} \right]=0.
\end{equation}
Taylor expansion of Eq.~(\ref{eq17}) around $r_s^\dagger$ to linear term and using Eq.~(\ref{eq16}) and combining with (\ref{eq18}) one obtains
 \begin{equation}\label{eq19}
r_s^*=r_s^\dagger-\frac{D_V(r_s^\dagger)}{\left[\frac{\partial D_N(r_s^\dagger)}{\partial r_s^\dagger}+\frac{\partial D_V(r_s^\dagger)}{\partial r_s^\dagger}\right]}.
 \end{equation}
 Since the energy $E/N$ is a convex function of $r_s$, both terms in the denominator of Eq.~(\ref{eq19}) are positive while the sign of the numerator is positive for $r_s^*<r_s^\dagger$ and negative for $r_s^*>r_s^\dagger$.

\end{document}